\documentclass{article}

\usepackage{PRIMEarxiv}

\usepackage[utf8]{inputenc} 
\usepackage[T1]{fontenc}    
\usepackage[hidelinks]{hyperref}       
\usepackage{url}            
\usepackage{booktabs}       
\usepackage{amsfonts}       
\usepackage{nicefrac}       
\usepackage{microtype}      
\usepackage{lipsum}
\usepackage{fancyhdr}       
\usepackage{graphicx}       
\graphicspath{{media/}}     
\usepackage{amsmath} 
\usepackage{algorithm} 
\usepackage{algpseudocode} 
\usepackage{amssymb} 
\usepackage{cleveref}
\usepackage{xcolor}
\usepackage{graphicx}
\usepackage{subcaption}
\graphicspath{ {./figures/} }
\title{Data-driven numerical site response
}

\author{
Joaquin Garcia-Suarez, Arthur Cornet, Sacha Wattel and  Jean-François Molinari \\
            Institute of Civil Engineering, Institute of Materials,\\
            \'{E}cole Polytechnique F\'{e}d\'{e}rale de Lausanne (EPFL), CH 1015 Lausanne, 
            Switzerland \\
  \texttt{\{joaquin.garciasuarez, arthur.cornet, sacha.wattel, jean-francois.molinari\}@epfl.ch} \\
}

\begin{document}
\maketitle

\begin{abstract}
Prediction of ground motion triggered by earthquakes is a prime concern for both the seismology community and geotechnical earthquake engineering one. The subfield occupied with such a problem is termed site response analysis (SRA), its one-dimensional flavor (1D-SRA) being particularly popular given its simplicity. 
Despite the simple geometrical setting, a paramount challenge remains when it comes to numerically consider intense shaking in the soft upper soil strata of the crust: how to mathematically model the high-strain, dissipative, potentially rate-dependent soil behavior. Both heuristics models and phenomenological constitutive laws have been developed to meet the challenge, neither of them being exempt of either numerical limitations or physical inadequacies or both. 
We propose herein to bring the novel data-driven paradigm to bear, thus giving away with the need to construct constitutive behavior models altogether. Data-driven computational mechanics (DDCM) is a novel paradigm in solid mechanics that is gaining popularity; in particular, the multiscale version of it relies on studying the response of the microstructure (in the case of soil, representative volumes containing grains) to populate a dataset that is later used to inform the response at the macroscale. This manuscript presents the first application of multiscale DDCM to 1D-SRA: first, we demonstrate its capacity to handle wave propagation problems using discrete datasets, obtained via sampling grain ensembles using the discrete element method (DEM), in lieu of a constitutive law and then we apply it specifically to analyze the propagation of harmonic waves in a soft soil deposit that overlies rigid bedrock. The soil in this application displays elastic yet non-linear response that is depth-dependent due to evolving overburden pressure. 
We validate the implementation via comparison to regular finite elements analyses (FEA), discuss the benefits of using DD, and demonstrate that traditional amplification functions are recovered when using the DDCM. Finally, we discuss a number of future work opportunities that lie ahead of this proof-of-concept, chiefly in terms of site-specific studies and incorporating more complex realistic traits of soil behavior. This project was developed using open-source software and the relevant code is freely made available to other researchers.
\end{abstract}


\section{Introduction}

Site response analysis aims at predicting the ground motion elicited by an earthquake, given certain information about the incoming seismic wavefront. In its simplest guise, we talk about 1D site response analysis (1D-SRA) \cite{Kramer}, as we assume full decoupling of the bulk waves (P, SV and SH) when they propagate perpendicularly to the ground surface and to every material interface in the soil layers, all of these being assumed to be perfectly horizontal.   

Of special importance is understanding the response of soft soil deposits that rest over much-stiffer rock, as this scenario entails the most destructive events \cite{Mexico}. 
The 1D assumption greatly simplifies the site response analyses (without compromising its scientific relevance \cite{Thompson-Baise}), but a major challenge remains: how to model the material response of the upper soil strata. In this setting, soil stiffness strongly depends, firstly, on depth as the confinement pressure increases the deeper the layer, and secondly, in the case of intense vibration, on the straining level, what renders a non-linear problem \cite{hardin}.

Linear-elastic 1D-SRA assumes that the soil only undergoes small strains, thus the problem can be framed in terms of elasticity theory, analytical solutions can be obtained \cite{Vrettos,ROVITHIS2011879}, and other mathematical tools as asymptotic analysis can be employed right away \cite{geotechnique_1,geotechnique_2}. On the other hand, constitutive modeling of the non-linear, dissipative, large-deformation and rate-dependent soil response has been and still is a subject of intense research \cite{Cam_clay,Whittle,SANISAND,Boulanger,UBCsand}.

The first efforts that tried to account for inelastic site response are attributed to Seed and Idriss \cite{Seed-Idriss}, who set up an scheme to iteratively recompute the elastic moduli and the dissipation for every strain level (while the former reduce, the latter increases with increasing straining). 
The so-called hyperbolic model was introduced around the same time to model cyclic sand response in the low to intermediate strain range \cite{hyperbolic}. 
In opposition to these heuristic approaches, researchers tried to appeal to plasticity theory to develop convenient constitutive models \cite{Cam_clay,Whittle}, the model parameters being calibrated using laboratory experiments. To give a sense of the number of parameters that require careful calibration, and of the number and breadth of the experiments to acquire data to do so, let us mention that the state-of-the-art model for sand plasticity requires over 20 independent parameters \cite{Boulanger}. 
It is therefore logical to look for alternatives to constitutive modeling \cite{Masi,Koreans_ML}. 
An obvious option would be to avoid homogenizing the grain ensemble altogether and resolving the microstructure (grains) in full detail using the discrete element method (DEM); however, this kind of simulations are computationally demanding. Moreover, one should take advtange of the fact that grain ensembles allow the proper definition of representative volume elements (``RVEs”) when enough grains are considered and subject to uniform deformation states at the RVE level \cite{DEM-book}.  

Data-driven computational mechanics (DDCM) \cite{Trent,Trent_2} was introduced, partially, as a way to circumvent the issues associated to the definition of complex phenomenological constitutive laws. The gist of it is to use data (that would be used anyways to calibrate a mathematical model of material behavior) directly in the computations. This opens a new paradigm of computational mechanics, in which the definition of convoluted material-response functions is avoided; instead, the traditional concepts of internal equilibrium and deformation compatibility are transferred to a “phase space” that material data inhabits; in this abstract mathematical space the solution of the problem amounts to matching the physically-admissible states to datapoints, for each and every element, by minimizing a certain phase-space distance between the points. The dataset that populates the phase space can come either from experiments \cite{LEYGUE} or from microstructure simulations \cite{Kostas_1}. More will be said on the latter when discussing applications to granular media.

\begin{figure}
    \centering
    \includegraphics[width=0.75\textwidth]{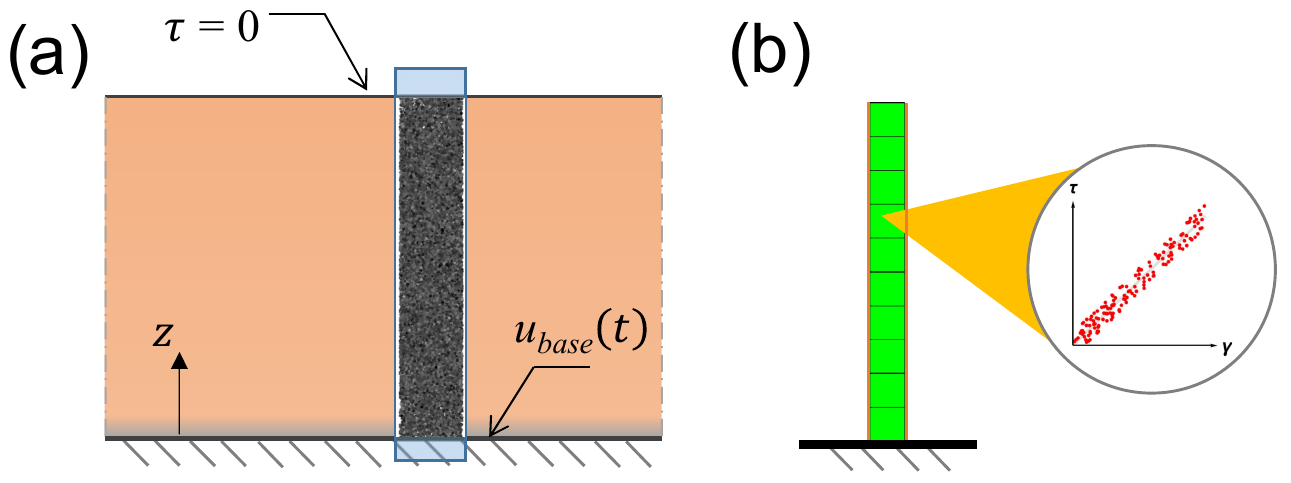}
    \caption{Scheme of the soil column model for site response analysis. (a) Soil layer on rigid bedrock homogeneized as a continuum, subject to horizontal base motion $u_{base}(t)$ (at $z=0$) and stress-free top surface ($\tau = 0$); the granular microstructure is revealed in the highlighted blue portion. (b) The just-mentioned highlighted soil column can be discretized using a mesh and a stress-strain ($\tau$-$\gamma$) dataset is then associated to each element to carry out a data-driven 1D site response analysis (orange flanks meant to represent periodic boundary conditions).}
    \label{fig:scheme}
\end{figure}


DDCM is a flourishing subject since it brings data considerations where they are needed the most (material response) without obfuscating the fundamental physics (i.e., equilibrium, compatibility), current efforts in the field range from optimizing the search space to improve numerical efficiency \cite{EGGERSMANN_2} to solving fracture mechanics problems \cite{Pietro}. Inelastic (dissipative) material behavior has already been considered in the context of DDCM, both in a general setting of internal variables \cite{eggersmann} and in terms of the dissipation inequality \cite{Kostas_1,Kostas_3}. 

DDCM appears ideally positioned to deal with problems that involve complex material response, granular media being a textbook example. Regarding constitutive modeling of granular media, Karapiperis and colleagues \cite{Kostas_1,Kostas_3} managed to reproduce experimental observations in meter-scale experiments by equipping FEM meshes with datasets mined from grain-scale simulations. Effectively, these were the pioneering multiscale data-driven simulations, which can be conceptualized as $FE^2$ simulations (e.g., \cite{Feyel}) in which microscale results are computed beforehand (instead of “on the fly”) and that, moreover, do not require definition of “tangent operators”, as the iterations of a, say, Newton-Raphson algorithm are replaced by phase-space searches. %
In addition, they also put forward an approach to deal with inelasticity that does not require internal variables but computes dissipation for each microscale ensemble and then forces the search algorithm to select thermodynamically-consistent points of ever-increasing dissipation in every element. 

The aforementioned applications could be considered the first DDCM applications to geotechnical engineering, but, to the best of our knowledge, no work has been done to bring DDCM to either seismic wave propagation or geotechnical earthquake engineering, let alone to 1D-SRA, a subject that can greatly benefit from it as it would circumvent the issues associated to approximating dynamic large-amplitude soil behavior with a mathematical function \cite{Boulanger}.

This text is intended to be a proof-of-concept showcasing the potential of DDCM to solve problems in 1D-SRA. First, we will demonstrate the capacity of DDCM to handle wave propagation in a continuum. Then, we will solve a 1D soil column (representative of a cross section of a soft soil deposit over bedrock) subjected to harmonic base shake, in which the material stiffness is mediated by the overburden pressure \cite{Kramer}. We will show that the bottom material displays linear elastic behavior while the upper layers behave elastically but non-linearly due to the lower confinement. Conveniently, we will be solving a conceptually-equivalent problem to the test model presented in Ref.~\cite{Trent_2}, just switching from a structural dynamics setting to continuum wave propagation. As the scope of this proof-of-concept is limited to non-dissipative behavior (to be tackled in the future), we must ensure small strains to guarantee negligible dissipation and elastic response. Future work will be tasked with including amplitude-dependent and frequency-dependent dissipation \cite{Domniki} as well as considering broadband earthquake motion \cite{Kramer}.



\section{Methods}

\subsection{Obtaining material data}
\label{Sec:DEM}

We wish to obtain data concerning homogenized microstructural response to then use it to inform the macrostructural response. 
LAMMPS (Large-scale Atomic/Molecular Massively Parallel Simulator) \cite{LAMMPS} is our tool of choice for the data mining process, due to its parallelization capabilities and robust DEM implementation. 

We are to use 2D grains with diameter either 2.5mm or 3.5mm and to assume that the material properties correspond to glass beads (using this material obeys to considerations of experimental convenience, as the spheres that are typically used in experiments are made of this material \cite{Ancey}), thus the grain density is fixed to $\rho=2500\, \mathrm{kg/m^3}$, the Poisson’s ratio to $\nu=0.3$ and the Young’s modulus to $E=50\,\mathrm{GPa}$ ($\mu = 19.2\,\mathrm{GPa}$). We choose to fix intergranular friction to zero $f=0$ (null intergranular friction coefficient) because the coming simulations will be restricted to the elastic regime where no dissipation happens. In passing, this also allows us to discriminate more easily the dissipation mechanisms present in this minimal 1D-SRA model. Thus, the LAMMPS input parameters are $k_t=0$ (null tangential contact stiffness, as friction is neglected) and $k_n = 36\, \mathrm{GPa}$ (normal contact stiffness). This translates into zero tangential force while the normal one comes given by

\begin{align}
\label{eq:DEM_force}
	\boldsymbol{F}_n 
    =
    \sqrt{ \delta R_{eq} }  
    \left(
    k_n \delta \boldsymbol{n}
    -
    \Gamma_n m_{eff} \boldsymbol{v}_n 
    \right)
    \, ,
\end{align}  

the first addend represents the Hertzian contact force, $R_{eq}$ being the harmonic mean of the touching particles’ radii, $\delta$ the intergranular penetration and $\boldsymbol{n}$ is the normal unit vector along the contact; the second addend represents a viscous contact force, proportional to the relative normal velocity between the grains $\boldsymbol{v}_n $, which is added for the sake of numerical stability \cite{Brilliantov}, $m_{eff}$ is the harmonic mean of the touching particles’ masses and we set the viscosity coefficient to $\Gamma_n = 0.2\, \mathrm{m^{-1} s^{-1}}$ \cite{behrooz}, the latter is included for numerical convenience  \cite{DEM-book}. Finally, let us mention that the timestep is picked based on the criterion $\Delta t < 0.14 \sqrt{m_{\text{min}} / k_{\text{max}}} $ \cite{otsubo_empirical_2017}, with $m_{\text{min}}$ being the minimum grain mass and $k_{\text{max}}$ the maximum normal contact stiffness. A thorough description of the particular data mining process is consigned to section \Cref{Sec:mining}.

\subsection{Data-driven computational mechanics}
\label{Sec:DD}

We are to show that the data-driven framework can handle time-domain wave propagation in a continuum, as it has not done before (to the best of our knowledge, only frequency-domain steady-state response has recently been tackled \cite{Amith}). The closest case is a structural dynamics study that was introduced during the presentation of the DDCM extension to dynamics \cite{Trent_2}.

Exactly the same DDCM algorithm can be used in the continuum case, adapting the meaning of the relevant variables from longitudinal wave propagation in rods to shear wave propagation in a continuum. Extensions to 2D and 3D wave propagation in continua may require extra tweaks to the basic flow utilized here; such an endeavor is part of our current work.    

Since we are presenting DDCM as a novelty in the field, let us restate the algorithm, in its distance-minimization search version (there is an alternative max-entropy search \cite{Trent_3}) as presented in Ref.~\cite{Trent}.

We use the fixed-point distance-minimization algorithm to match a physically-feasible state, in terms of both compatibility and equilibrium, to the right data point. The elements are initialized to a random material solution set, and then iterations follow: find the closest admissible solution to the current guess, then find the closest material solution set to this admissible solution. The algorithm is said to have converged when two consecutive iterations yield the same material point selection. 

In the dynamic framework \cite{Trent_2}, the method, for a system made up of $N_e$ elements and $N_n$ nodes,
is the following: given an element $e$ in a state $(\sigma_e^{*(k)},\varepsilon_e^{*(k)})$ (recall $\gamma_e = 2\varepsilon_e$) from the dataset, $k$ being the iteration index, we compute the closest ``admissible state'' $(\sigma_e^{(k)},\varepsilon_e^{(k)})$ as:

\begin{equation}
\label{eq:epsi_sigma}
    \begin{cases}
      \varepsilon_e^{(k)} = B_e u^{(k)}
      \vspace{5pt} 
      \\
      \sigma_e^{(k)} = \sigma_e^{*(k)} + \mathbb{C} B_e \eta^{(k)}
    \end{cases} \, ,
\end{equation}

where $B_e$ relates displacements to strains \cite{Trent}, $\mathbb{C}$ is a numerical constant,
$u^{(k)} \in \mathbb{R}^{N_n \times 1}$ is a column vector that contains the horizontal displacement of each node and $\eta^{(k)} \in \mathbb{R}^{N_n \times 1}$ are the Lagrange multipliers (enforcing the nodal equilibrium constraint) at the $k$-th iteration, these are established using the following equation, minimizing a certain functional \cite{Trent_2}:

\begin{equation}
\label{eq:system_full}
    \begin{cases}
      \left(\sum_{e} w_e B_e^\top \mathbb{C} B_e\right) u^{(k)} = \sum_{e} w_e B_e^\top \mathbb{C} \varepsilon_e^{*(k)} + \dfrac{M}{\beta \Delta t^2} \eta^{(k)} \vspace{5pt} 
      \\
      \left(\sum_{e} w_e B_e^\top \mathbb{C} B_e\right) \eta^{(k)} = f - \sum_{e} w_e B_e^\top \sigma_e^{*(k)} - \dfrac{M}{\beta \Delta t^2} (u^{(k)} - u^{pred})
    \end{cases} \, ,
\end{equation}
$w_e$ is a per-element summation weight,
$M \in \mathbb{R}^{N_n \times N_n}$ is the nodal mass matrix, $\Delta t$ is the timestep and $\beta$ is a Newmark method's constant (see below).
If we define the two matrices $K_{eq} = \sum_{e} w_e B_e^\top \mathbb{C} B_e$ and $\Lambda = \frac{M}{\beta \Delta t^2}$, equation~\ref{eq:system_full} can be rewritten as the following system:
\begin{equation}
\label{eq:system_matrix}
    \begin{bmatrix}
    K_{eq} & - \Lambda\\
    \Lambda & K_{eq}
    \end{bmatrix}
    \begin{bmatrix}
    u^{(k)}\\
    \eta^{(k)}
    \end{bmatrix}
    = 
    \begin{bmatrix}
    \sum_{e} w_e B_e^\top \mathbb{C} \varepsilon_e^{*(k)}\\
    f - \sum_{e} w_e B_e^\top \sigma_e^{*(k)} + \Lambda u^{pred}
    \end{bmatrix} \, .
\end{equation}

The Newmark method \cite{Cook} has been used to integrate the equations in time. At each time step $n$, the velocity $v_n$ and acceleration $a_n$ at each node are computed using:
\begin{equation}
\label{eq:Newmark}
    \begin{cases}
      a_n = \dfrac{u_n - u_n^{pred}}{\beta \Delta t^2}
      \vspace{5pt} 
      \\
      v_n = v_n^{pred} + \gamma a_n \Delta t
    \end{cases} \, ,
\end{equation}
where $\beta = 0.25$ and $\gamma = 0.5$ are the method's parameter values. Finally, the predicted displacements $u_n^{pred}$ and velocity $v_n^{pred}$ are computed as follows:

\begin{equation}
\label{eq:Newmark_prediction}
    \begin{cases}
      u_n^{pred} = u_{n-1} + v_{n-1} \Delta t + \left( \frac{1}{2} - \beta \right) a_{n-1} \Delta t^2
      \vspace{5pt} 
      \\
      v_n^{pred} = v_{n-1} + (1-\gamma)a_{n-1} \Delta t
    \end{cases} \, .
\end{equation}

Using all these equations, we have therefore determined the closest admissible state $(\sigma_e^{(k)},\varepsilon_e^{(k)})$. We can then select the closest state $(\sigma_e^{*(k+1)},\varepsilon_e^{*(k+1)})$ from the dataset and start over until we reach convergence. Further details as to numerical aspects of the method can be found in Refs.~\cite{Trent,Trent_2,EGGERSMANN_2}.

\begin{algorithm}[H]
\caption{Fixed-point algorithm for dynamics}\label{alg:FDPM}
\begin{algorithmic}

\State Define strain-deformation matrix $B_e$, mass matrix $M$ and matrix $\mathbb{C}$ containing distance constants.
\State Choose Newmark parameters $\beta$ and $\gamma$, time step $\Delta t$, and initialize $a_0$ and $v_0$
\State Compute $K_{eq} = \sum_{e} w_e B_e^\top \mathbb{C} B_e$ and $\Lambda = \frac{M}{\beta \Delta t^2}$
\State Choose $(\sigma_e^{*(0)}(0),\varepsilon_e^{*(0)}(0))$ randomly from dataset.

\For{n $\leq$ number of time steps ($\Delta t$)}
\State $(\sigma_e^{*(0)}(n),\varepsilon_e^{*(0)}(n)) \gets (\sigma_e^{*(0)}(n-1),\varepsilon_e^{*(0)}(n-1))$
\State Compute predictions $u_n^{pred}$ and $v_n^{pred}$ (eq.~\ref{eq:Newmark_prediction})
\State Compute $a_n$ and $v_n$ (eq.~\ref{eq:Newmark})
\State $k \gets 0$

\While{$(\sigma_e^{*(k)},\varepsilon_e^{*(k)}) \neq (\sigma_e^{*(k-1)},\varepsilon_e^{*(k-1)})$}
\State $k \gets k+1$
\State Compute $u^{(k)}$ and $\eta^{(k)}$ (eq.~\ref{eq:system_matrix})
\State Compute corresponding admissible state $(\sigma_e^{(k)},\varepsilon_e^{(k)})$ (eq.~\ref{eq:epsi_sigma})
\State Find closest state $(\sigma_e^{*(k)},\varepsilon_e^{*(k)})$ from data set
\EndWhile

\State Return $(\sigma_e^{*}(n),\varepsilon_e^{*}(n))$
\EndFor

\end{algorithmic}
\end{algorithm}

%
\section{Proof-of-concept simulations}

\subsection{Data-driven wave propagation}
\label{Sec:DD-FEM}
We consider the case of shear wave propagation in a rope-like setting \cite{Graff}, total length $L$, in which one end is whipped imposing a perpendicular displacement $u_{base}(t)=u_0 \sin \left( 2\pi t / T \right)$ for $0<t<T$ and $u=0$ otherwise ($T = 5 \, \mathrm{s}$ being the oscillation period) and fixed at the other end. The PDE to solve is

\begin{align} \label{eq:PDE}
    {\partial \over \partial z} 
    \left[ 
    \mu(z) 
    {\partial u \over \partial z} 
    \right]
    =
    {\partial^2 u \over \partial t^2} \, ,
\end{align}

subject to quiescent initial conditions $u(z,t=0) = \dot{u}(z,t=0) = 0$, and boundary conditions $u(z=0,t)=u_{base}(t)$ at the bottom and $u(z=L,t)=0$ at the top.

\begin{figure}
\centering
\captionsetup[subfigure]{justification=centering}
\begin{subfigure}[b]{.45\linewidth}
\includegraphics[width=\linewidth]{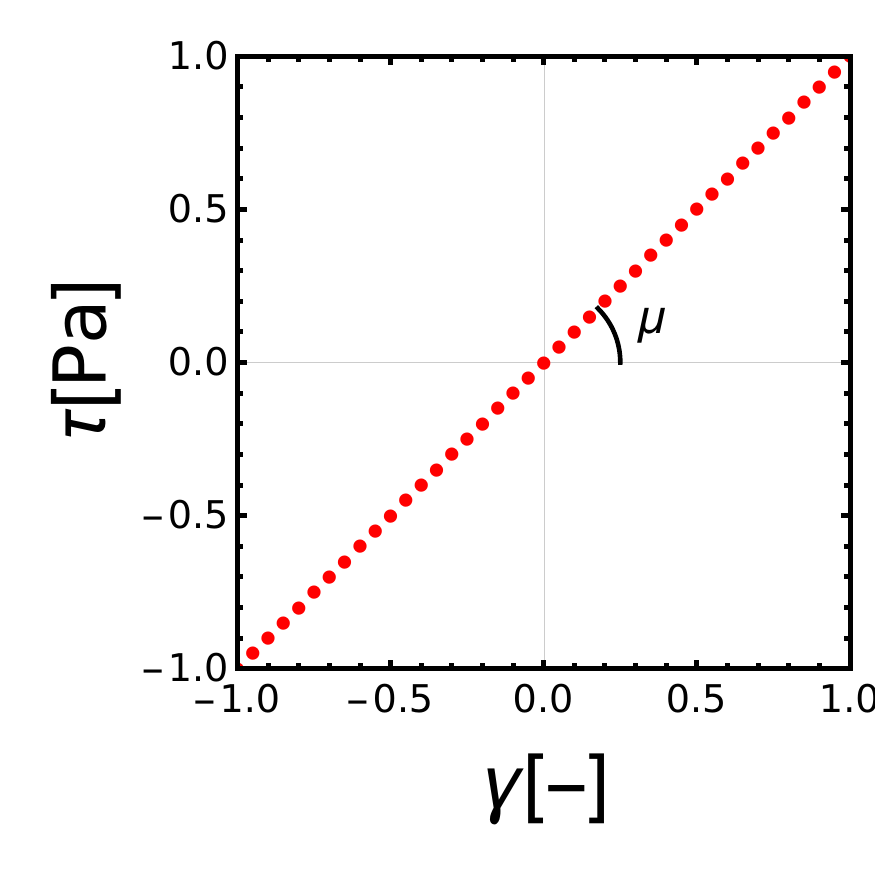}
\caption{Dataset used for the wave propagation example.}\label{fig:rope_dataset}
\end{subfigure}
\begin{subfigure}[b]{.45\linewidth}
\includegraphics[width=\linewidth]{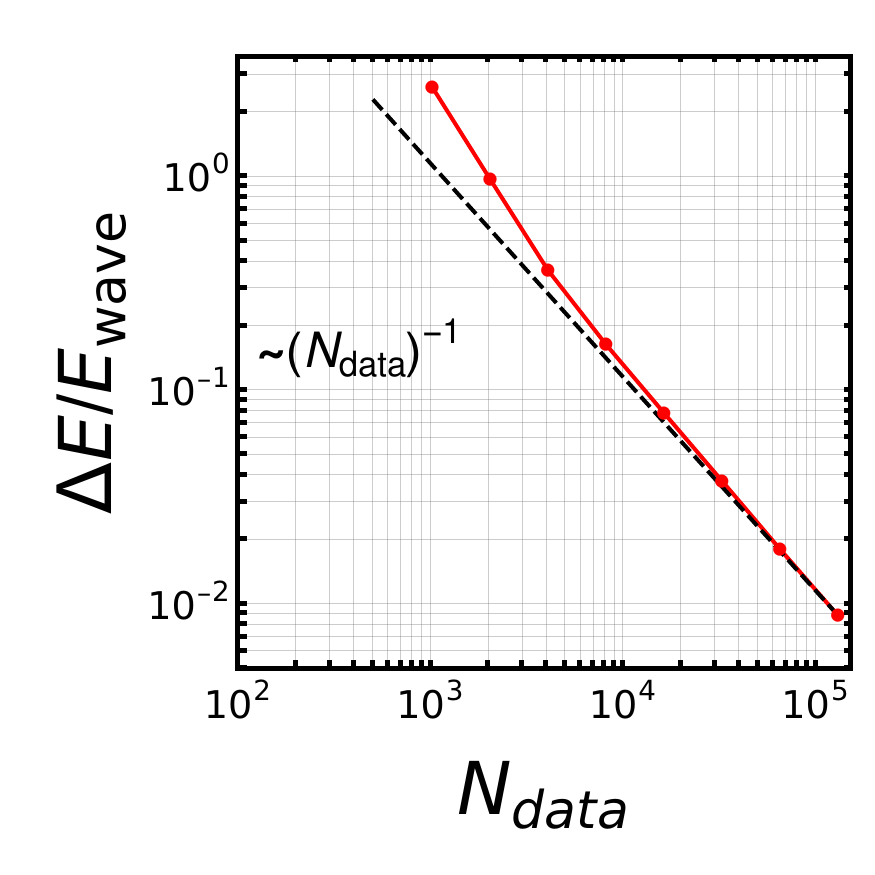}
\caption{Energy scaling with $N_{data}$.}\label{fig:rope_energy_scaling}
\end{subfigure}
\begin{subfigure}[b]{.45\linewidth}
\includegraphics[width=\linewidth]{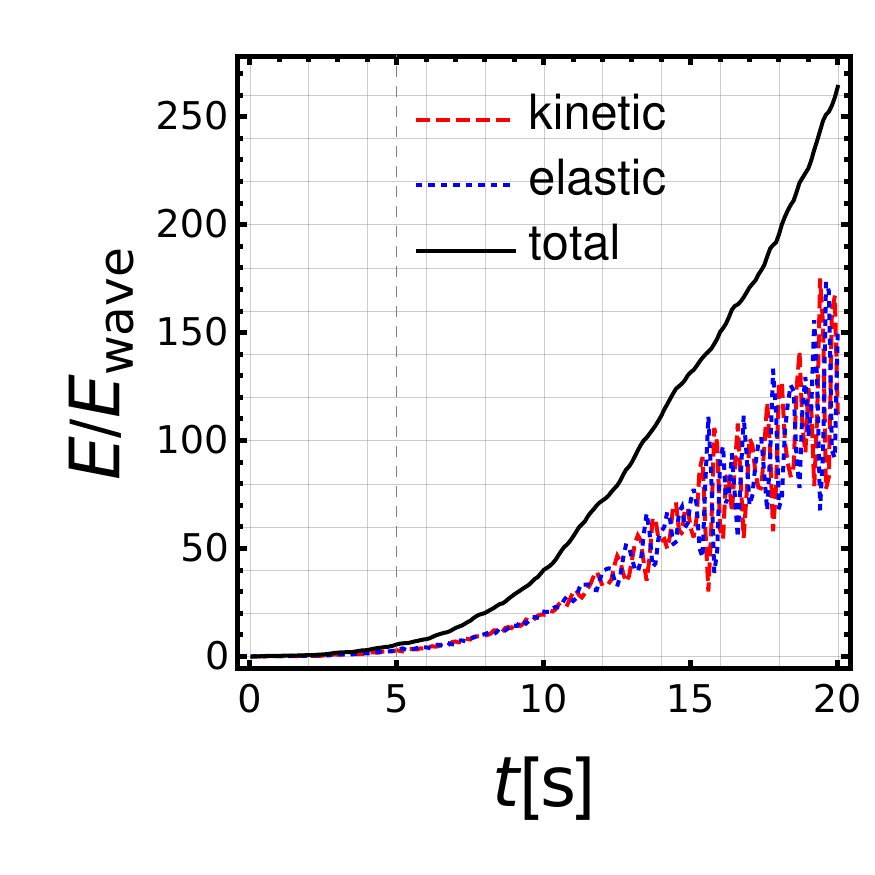}
\caption{Energy tracking during simulation\\ (small dataset $N_{data} = 101$).}\label{fig:rope_small_dataset}
\end{subfigure}
\begin{subfigure}[b]{.45\linewidth}
\bigskip
\includegraphics[width=\linewidth]{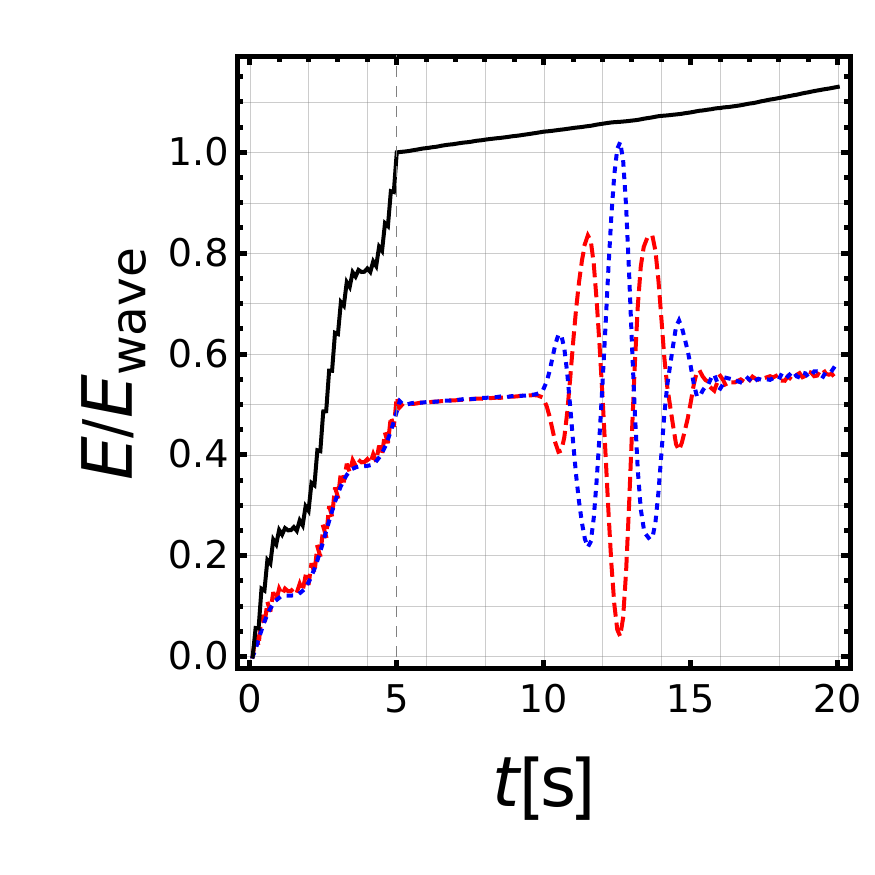}
\caption{Energy tracking during simulation\\ (large dataset $N_{data} = 10001$).}\label{fig:rope_large_dataset}
\end{subfigure}
\caption{Wave propagation example. (a) Material response dataset (the same for every element) and underlying shear modulus $\mu$. (b) Trend over multiple simulations with varying $N_{data}$: the usual scaling $\sim N_{data}^{-1}$ \cite{Trent_2} is recovered. (c) Extra energy in the model over simulation duration: clear artificial energy creation after the imposed displacement ends: final energy is over 250x the total energy transported by the wave ($E_{wave}$). (d) Extra energy in the model during simulation duration: small-yet-noticeable artificial energy creation after the imposed displacement ends (extra $\sim 10\%$ of correct value $E_{wave}$), the later large oscillations correspond to the wave interaction with the fixed end during the reflection process.}
\label{fig:rope}
\end{figure}

We choose test values for this simulation: the total length is $L = 10\, \mathrm{m}$ and the density is $\rho = 1 \, \mathrm{kg/m^3}$  and the underlying shear modulus is taken to be $\mu = 1 \, \mathrm{Pa}$. We specify “underlying” because the DD solver will not be aware of its value, it will only match physical values to the phase space point contained in a dataset that is in turn sampled from the “underlying” linear-elastic constitutive law $\tau = \mu \gamma$. The sampling (\Cref{fig:rope_dataset}) is done uniformly in strain (y-axis), so increasing the sampling frequency amounts to a larger dataset. We illustrate the effect of dataset size and will reproduce some results as to error scaling that have been reported in the literature \cite{Trent}, \Cref{fig:rope_energy_scaling}. The time integration is carried out using Newmark's method with $\beta = 0.25$ and $\gamma = 0.5$ (see \Cref{Sec:DD}). The simulation runs for a total time of $20 \, \mathrm{s}$ with timestep $\Delta t =0.1 \, \mathrm{s}$.

\begin{figure}
    \centering
    \includegraphics[width=0.8\textwidth]{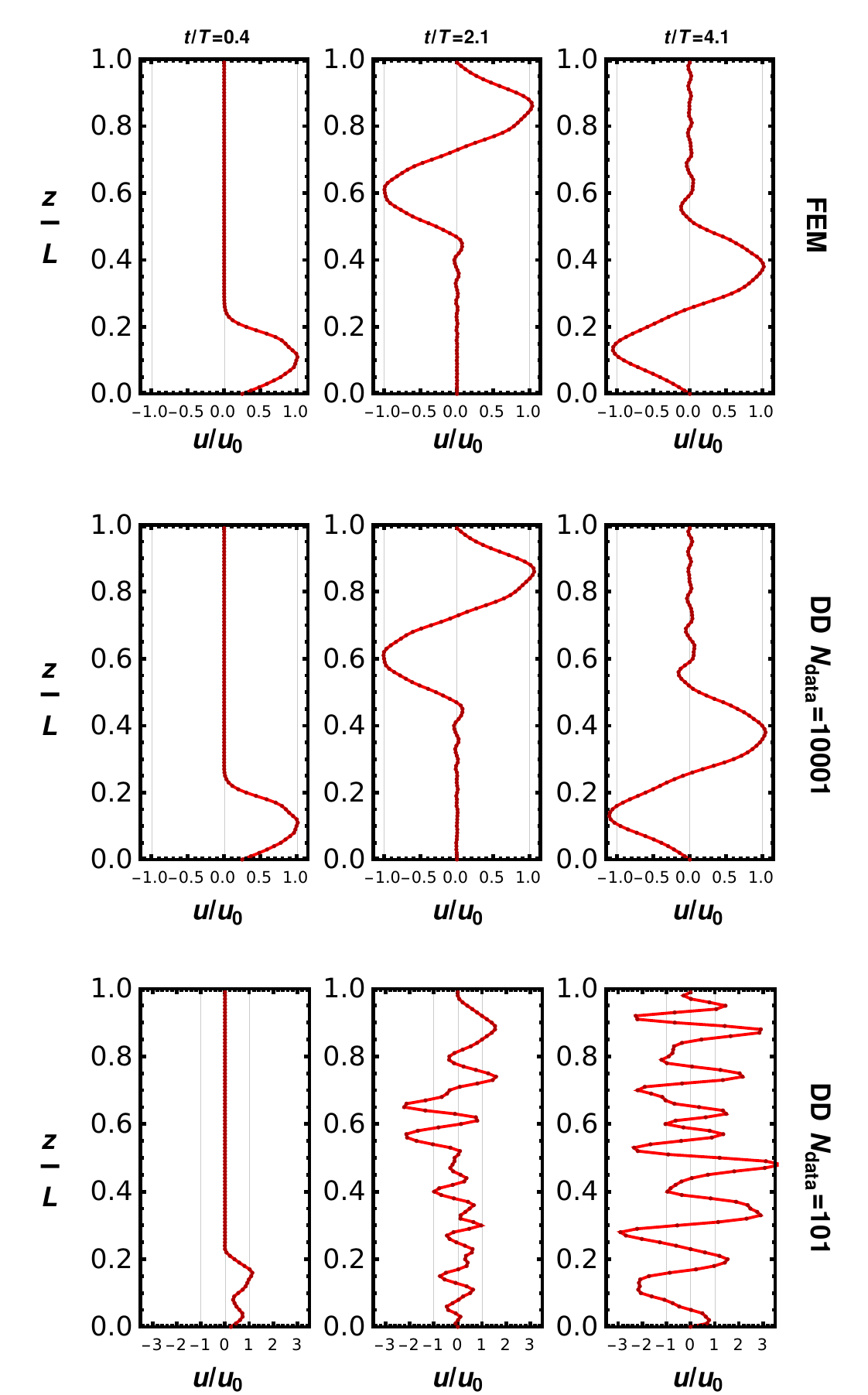}
    \caption{Data-driven 1D wave propagation: each column corresponds to a time snapshot while each row is one of three solvers (from top to bottom: traditional FEM, DD with rich dataset, DD with poor dataset). See that the two first rows are virtually indistinguishable while the bottom one is clearly unable to propagate the wave (notice the different horizontal range in these plots).}
    \label{fig:rope_wave}
\end{figure}

The simulation results displayed in \Cref{fig:rope_wave} reveal the ability of the DD framework to propagate the wave as satisfactorily as the FEM when enough datapoints are considered ($N_{data}$ being the number of points in the set): 
the smaller set ($N_{data}=101$ points, bottom row in \Cref{fig:rope}) displays nonphysical noisy results while the larger one ($N_{data}=10001$ points, middle row) is indistinguishable from the FEM solution (top row).

We compute the total energy (straining plus kinetic) that is present in the model, \Cref{fig:rope_small_dataset,fig:rope_large_dataset}, and acknowledge that, when the dataset is small \Cref{fig:rope_small_dataset}, substantial amounts of energy are artificially introduced by the DDCM procedure as the distance between physically-admissible points in phase space can be sizeable. The scaling observed for this “excess energy” $\Delta E$ is $\sim N_{data}^{-1}$ when the dataset is large enough, \Cref{fig:rope_energy_scaling}, in agreement to the trends reported in the literature previously for noise-less datasets\cite{Trent}.


\subsection{Seismic response soil deposit}

\subsubsection{Data-mining and model set up}
\label{Sec:mining}

\begin{figure}
    \centering
    \includegraphics[width=\textwidth]{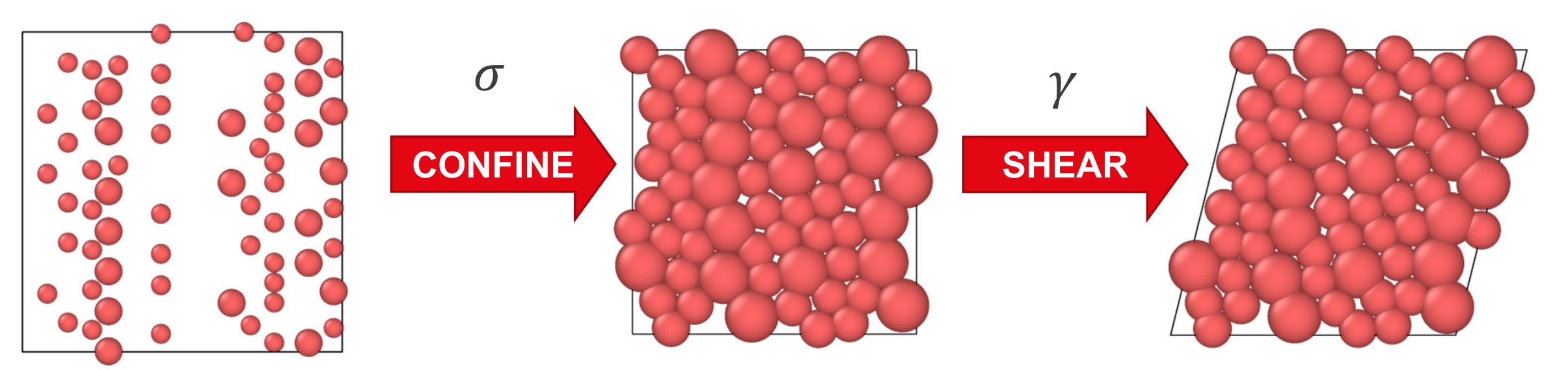}
    \caption{Probing grain ensemble response using DEM: from left to right, random particle insertion, confinement up to desired level $\sigma$, imposing box angular distorion $\gamma$ (shearing). Note: the actual boxes contain more grains.}
    \label{fig:process}
\end{figure}

For the subsequent DDCM simulations, we require a dataset parametrized by the following phase space coordinates: $\tau$ (shear stress), $\gamma$ (shear strain i.e., angular distortion) and $\sigma$ (overburden pressure). Since $\sigma$ will remain unchanged for each element once its relative depth is specified, this variable can be taken as a “label” that permits assigning a different dataset to each element depending on their vertical position, thus reducing the phase space to a 2D plane $(\gamma, \, \tau)$. The steps leading to the microscale dataset creation are:

\begin{enumerate}
    \item Based on the height of the soil column and the foreseeable discretization thereof, choose adequate values of $\sigma$ to apply to the grain ensemble. There is a single $\sigma$ value for each element, corresponding to the pressure at the middle point thereof.
    \item LAMMPS initializes the particle ensemble inserting randomly the two possible sizes so that they do not contact at first (\Cref{fig:process}, left). Then, the box size is progressively reduced to bring the particles into contact (\Cref{fig:process}, center). As all boundaries are periodic, the pressure also rises. Once the desired level of $\sigma$ is attained, the box size is fixed and the system is left to relax until the excess kinetic energy dissipates. This process is iterated until the desired $\sigma$ is obtained after relaxation.
    Based on the number of particles that is known to be necessary to homogenize the elastic response of the ensemble \cite{DEM-book} and after our own verification, we set a total of 240 particles filling a $5\, \mathrm{cm} \times 5\, \mathrm{cm}$ box, what leads to a packing ratio equal to 0.8. 
    \item Keeping this hydrostatic pressure state, a final shear strain $\gamma$ is incrementally imposed by applying a uniform quasi-static displacement at the top edge while the bottom is kept fixed. The values of stress $\tau$ induced at each increment of $\gamma$ are recorded to a file labeled with the corresponding value of confinement $\sigma$. The file contains pairs $(\gamma,\, \tau)$ to be passed later to the DD solver.   
\end{enumerate} 

The confinement pressures are increased from $10\, \mathrm{kPa}$ to $200\, \mathrm{kPa}$ in increments of $10\, \mathrm{kPa}$. This makes for a total of 20 datasets, one per element in the discretized column.

Reproducing this data mining scheme requires LAMMPS (free open-source software) and a short Python code that can be readily downloaded from the second author Gitlab page, see Supplementary Material section.    


\begin{figure}
    \centering
    \includegraphics[width=\textwidth]{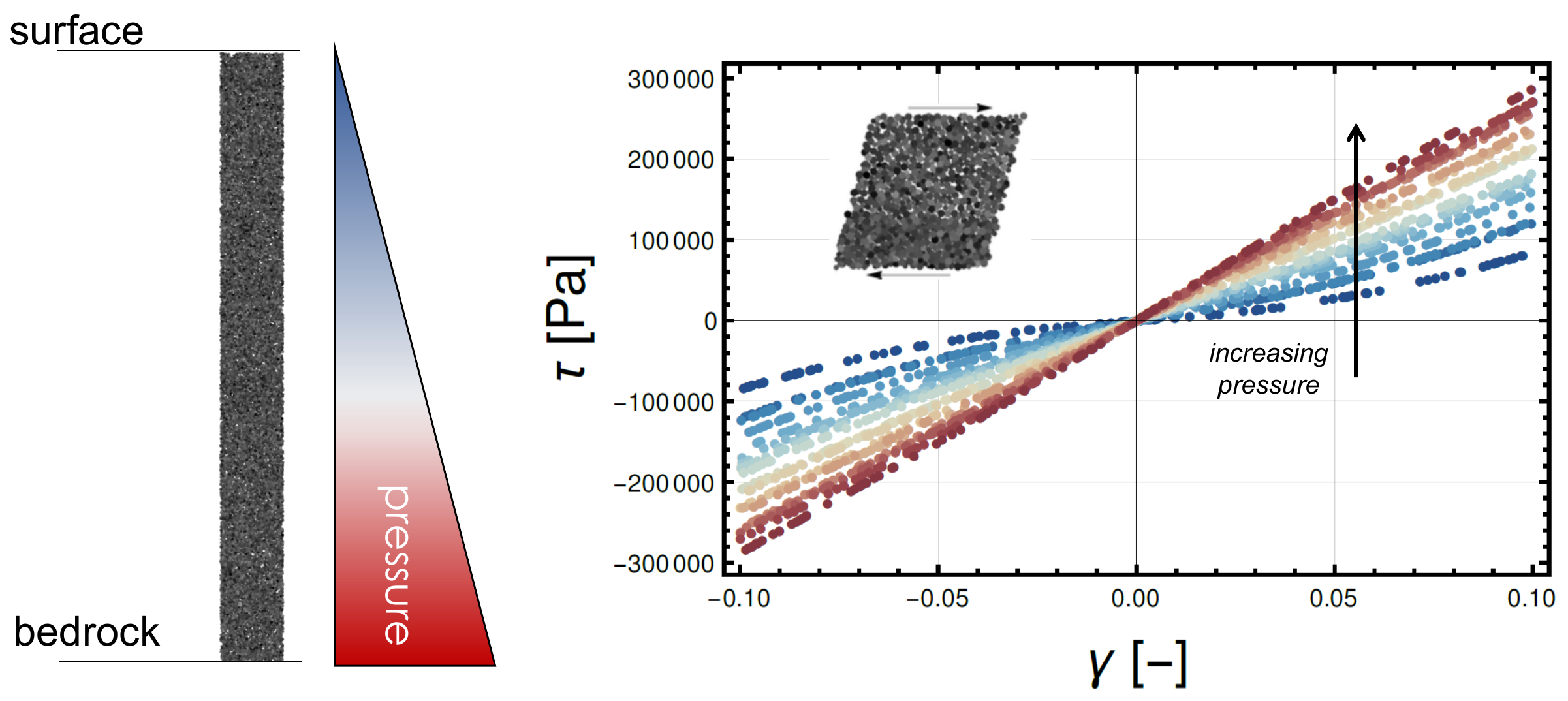}
    \caption{Depth-dependent response: Left, scheme of the soil column and pressure diagram; right, simultaneous visualization of the datasets (color code used for pressure is the same in both figures, notice the linear response in red while non-linear in dark blue). Datasets in display are smaller than those used in simulations.}
    \label{fig:material_response_SRA}
\end{figure}



\subsubsection{Soil deposit base shaking: comparison to FEM} \label{Sec:TFs}

Let us now consider an idealized soil column subject to harmonic horizontal base motion ($u$ being the horizontal displacement), initially quiescent (i.e., $u=0$ at $t=0$ for $0<z<H$). The imposed harmonic base motion comes defined by $u_{base} (t)= u_0 \sin (2\pi t/T)$, where the amplitude is to be chosen to guarantee small strains ($u_0 = 0.01 \, \mathrm{m}$) and the influence of the oscillation period $T$ will later be studied parametrically.

We do not focus on a particular site in the field but aim to model a possible laboratory setup: the height is chosen to be $H = 1\, \mathrm{m}$ and discretized in 20 linear elements of height $h =5\, \mathrm{cm}$. We choose this size as the box used to probe the grain ensemble response is $5\, \mathrm{cm} \times 5\, \mathrm{cm}$, this yields a system that could be realized experimentally with relative ease and, moreover, that could be fully resolved with DEM at an affordable computational cost. The grain material properties are the ones of commercial glass spherical beads (\Cref{Sec:DEM}). We use adjusted values of gravity to obtain a weight-induced overburden pressure consistent with the values used during the dataset creation ($10 \, \mathrm{kPa}, \, 20 \, \mathrm{kPa}, \, \ldots , 200 \, \mathrm{kPa}$). The dimensionless pressure number \cite{agnolin_internal_2007-1}, 
\begin{align} \label{eq:kappa}
    \kappa
    =
    \left( 
    {E^* \over \sigma}
    \right)^{3/2} \, ,
\end{align}
represents the ratio between grain stiffness and overburden pressure ($E^* = E / (1 - \nu^2)$, where note that the Young's modulus $E$ and the Poisson ratio $\nu$ are not the soil's but the grains'; a soil constitutive law constant is not necessary when performing DDCM computations), and relates to the overlapping between grains, $\delta$: $\delta / R_{eq} \sim \kappa^{-1}$ \cite{agnolin_internal_2007-1}. This number is useful when it comes to differentiate between different grain ensemble regimes: when $\kappa \gg 1$, there is little grain deformation, when $\kappa \sim 1$ substantial grain deformation, plasticity and potential breakage can take place. The values range from $\kappa \approx 1.25 \cdot 10^8$ at the topmost element to $\kappa \approx 1.11 \cdot 10^{10}$ at the bottom element, what means that in all cases the grains arrange into a ``rigid sphere packing'' that can deform elastically \cite{agnolin_internal_2007-1}.      

\begin{figure}
    \centering
    \includegraphics[width=\textwidth]{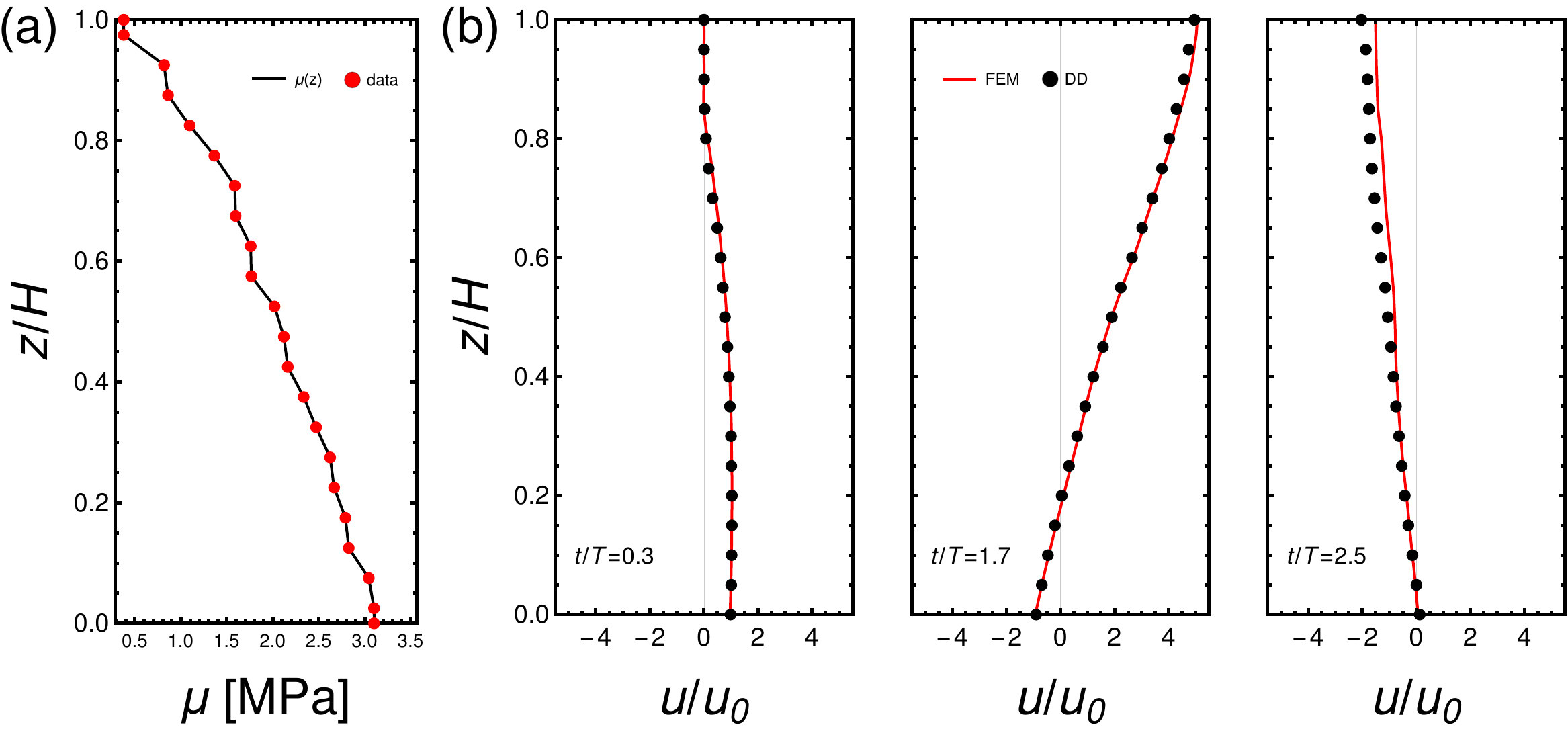}
    \caption{Comparison FEM-DD. (a) Shear modulus depth-dependence: points obtained from LAMMPS probing, and linear piece-wise interpolation passed to the FEM solver (the evolution from the extreme points to either the bottom or top values is assumed to be constant). (b) Three snapshots of wave propagation elicited by base harmonic shake (note divergence between continuous line and points in top part of rightmost panel).}
    \label{fig:DD-FEM}
\end{figure}

We set a FE model up in the traditional way for comparison purposes. The first step is to convert the datapoints in a function; if we were to define the whole range of observed behavior we would have to define a function of both depth ($z$ or equivalently $\sigma$) and strain level ($\gamma$) to account for the non-linear strain-dependent stiffness of the top layers (i.e., $\mu_{\text{FEM}} = \mu_{\text{FEM}}(\gamma,z)$), but, more realistically, the first choice would be to define a function of depth fixing a certain strain level, which would naturally mean no errors in the lower strata that do tend to behave more linearly while potentially some errors at the upper non-linear layers. We do the latter, extracting the values at $\gamma = 5 \%$ and defining a piece-wise linear interpolation function among the points $\mu_{\text{FEM}}= \mu_{\text{FEM}}(0.05,z) =\mu_{\text{FEM}} (z)$, \Cref{fig:DD-FEM}a. The FEM problem is solved in Mathematica \cite{Mathematica} using the NDSolveValue function with default parameters (in particular, the discretization of the linear domain is automatically selected, the code used for this FEM solution is being provided, see Supplementary Material).

The results for $u_0 = 0.01\, \mathrm{m}$  and $T = 0.1 \, \mathrm{s}$ are displayed in \Cref{fig:DD-FEM}. We observe that the agreement is reasonable except occasionally at the upper portion, see \Cref{fig:DD-FEM}(b) rightmost panel, what was to be expected as the FE model cannot handle the non-linear soil response there. More on this matter in \Cref{Sec:Discussion}.

\subsubsection{Base-to-top displacement transfer function: comparison to analytical solutions}

Once the DD 1D-SRA model has been verified, we can move to use it to construct transfer functions. These are frequency-domain relations between the displacement amplitude either at a rock outcrop or at a certain depth (in the case of soft soil deposits, at the bedrock $\hat{u}_{base}$) and at the free surface ($\hat{u}_{top}$). We focus on the base-to-top transfer function in a deposit overlying rigid bedrock, originally quiescent, imposing a small-amplitude sinusoidal displacement at the bottom starting at $t=0$ and recording the amplitude at the surface once steady-state conditions are attained. Repeating the simulation changing the period of the imposed displacement $T$ we sample various points of the transfer function. Here, instead of comparing to FEM results, we resort to analytical transfer functions \cite{Gazetas}, assuming that the datapoints were interpolated using a linear fit:

\begin{align}
	\mu(z) = \mu_{base} \left( 1 - \alpha \frac{z}{H} \right)
    \label{eq:fun2}
\end{align}
where $\alpha= \mu_{top} / \mu_{base} - 1$. Having the parameters that define $\mu =\mu(z)$, the base-to-top transfer function requires solving the frequency-domain version of \cref{eq:PDE} \cite{geotechnique_2} with boundary conditions $[\partial u / \partial z]_{z=L} = 0$ $\forall t$ at the top (equivalent to stress-free top surface condition, since $\mu(H) \ne 0$) and forced harmonic loading at the bottom; thus the transfer function $A$ comes given as \cite{Gazetas} 
\begin{align}
	A(\alpha,r) 
	=
	{\hat{u}_{top} \over \hat{u}_{base}}
	= 
	\frac{\alpha }{2 \sqrt{1-\alpha } \sqrt{-r^2} \left(J_0\left(\frac{2 r}{\alpha }\right) K_1\left(\frac{2 \sqrt{-r^2} \sqrt{1-\alpha }}{\alpha
   }\right)+I_1\left(\frac{2 \sqrt{-r^2} \sqrt{1-\alpha }}{\alpha }\right) K_0\left(\frac{2 \sqrt{-r^2}}{\alpha }\right)\right)} \, ,
    \label{eq:A_mylo}
\end{align}

where 
$r = \omega H / V_{base}$ ($\omega = 2\pi / T$ and $ V_{base} = \sqrt{\mu(z=0) / \rho}$), 
$J_0(x)$ is the Bessel function of the first kind and order zero,
$I_1(x)$ is the modified Bessel function of the first kind and order one and $K_0(x)$ is the modified Bessel function of the second kind and order zero, while $K_1(x)$ is order one. 
Hysteretic damping in frequency domain is introduced into the analytical transfer function via a complex shear modulus $\mu^*(z)= \mu  (1 + \mathrm{i} \delta_d)$ in which the value of the damping coefficient $\delta_d = 0.07$ will be chosen to match the amplitude of the fundamental resonance peak inferred from the DD results (even though this dissipation mechanism is distinct from the one in the multiscale model, i.e., viscous dissipation at the granular contacts, see \cref{eq:DEM_force}). 

The model allows obtaining a closed-form expression for the base-to-top transfer function, but only at the expense of important simplifications: on one hand, the material data can not be considered in its entirety, a reference level of strain has to be fixed and then interpolated; on the other one, a simplistic linear interpolation has to be used to obtain a version of \cref{eq:PDE} that can be solved analytically in the frequency domain.

\begin{figure}
    \centering
    \includegraphics[width=\textwidth]{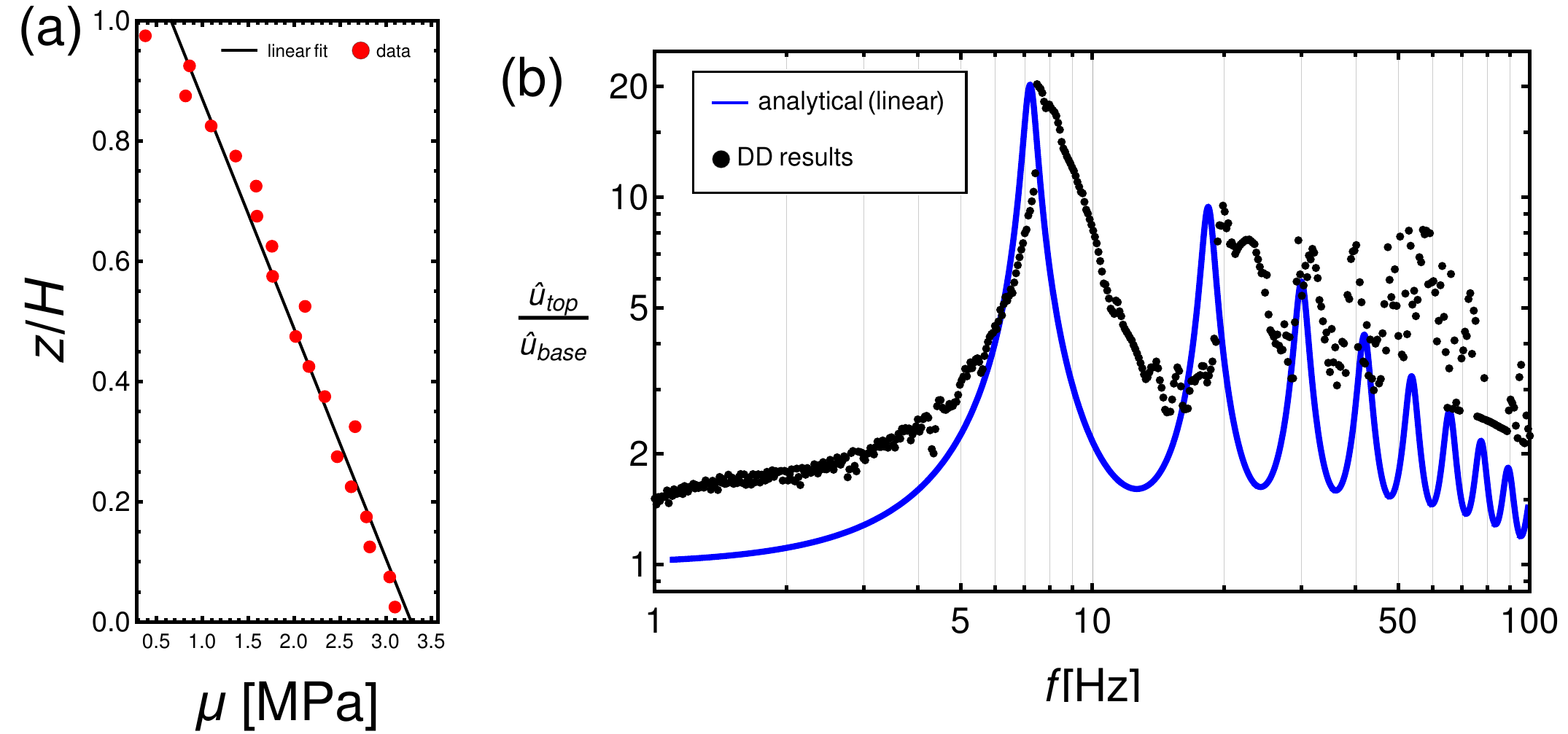}
    \caption{(a) Material stiffness data in depth at $\gamma = 1 \%$ and linear interpolation (minimal square regression). (b) Base-to-top displacement transfer function comparison: solid line represents the analytical transfer function \cref{eq:A_mylo} and points are the DD amplification results. The damping in the analytical model is chosen to be $\delta_d = 0.07$ in order to match the amplitude of the first resonance peak.}
    \label{fig:TF}
\end{figure}

In \Cref{fig:TF}b, we observe the qualitative resemblance between the analytical transfer function and the one inferred from the DD results: both of them display low amplification in the low frequencies (but the analytical model delivers a quicker ramping-up as the first resonance approaches), maximum amplitudes around a fundamental resonance peak, and oscillation amplitudes of less magnitude in the high-frequency portion. Quantitatively, the first resonance peak amplitude is matched by design (choosing $\delta_d = 0.07$), while the fundamental resonance frequencies match organically  ($\approx 7 \, \mathrm{Hz}$ in both cases). As to the rest of the frequency range, the DD model predicts greater amplitudes both in the low-frequency regime and high-frequency one, the location of the second resonance peak seems to match in both cases as well.

\section{Discussion}
\label{Sec:Discussion}

In \Cref{fig:TF}b, the fact that the analytical fundamental frequency matches the first peak inferred from the DD results means that the material stiffness at the lower layers is properly reflected \cite{Fundamental_mode,Mylonakis_exponential}, while the greater amplitudes in the low frequency regime could be due to FEA's lack of precision in accounting for the upper layers' stiffness and their non-linear behavior.

Remarkably, one of the main drawbacks of analytical models that assume constant damping irrespectively of loading amplitude is that they yield unrealistically-low amplitudes in the high-frequency range, which is not the case in the DD model: clearly, the amplitude remains substantial after the second resonance peak, whereas the analytical model predicts a fast decay.

Regarding the higher frequencies, one should also ask: can a model that does not consider rate effects yield useful results in the high-frequency range? To answer such a question we resort to another important dimensionless group in the study of granular media, the inertial number \cite{DEM-book}, $I$, that in 2D comes given by

\begin{align}
    I
    =
    \dot{\gamma}
    \sqrt{\rho d^2 \over \sigma} \, ,
\end{align}

where $d$ represents the average grain diameter ($3\, \mathrm{mm}$ in our case). $I \sim 1$ means that inertial effects at the grain scale can not be ignored, so, assuming a harmonic shear deformation, frequency $\mathsf{f}$, that induces strain of amplitude $u_0 / h$,

\begin{align}
    I
    \sim
    { 2 \pi \mathsf{f} u_0  
    \over
    h}
    \sqrt{ \rho d^2 \over \sigma} 
    \sim
    1
    \Rightarrow
    \mathsf{f}_{\text{lim}}
    \sim
    { h \over u_0}
    \sqrt{\sigma  \over \rho d^2 } \, ,
\end{align}

$\mathsf{f}_{\text{lim}}$ would be the shake frequency that delivers $I \sim 1$, fixing all the other values to the ones used in the study. Choosing the most unfavorable value of confinement ($\sigma = 20\, \mathrm{kPa}$) yields $\mathsf{f}_{\text{lim}} \sim 1000 \, \mathrm{Hz}$. Hence, given the parameters of this idealized model, it can be tested up to $100 \, \mathrm{Hz}$ without accounting for rate-dependent material response. Notwithstanding, we foresee the inertial number \cite{agnolin_internal_2007-1} being another coordinate we will use to parametrize the phase space along with the already-mentioned pressure number, \cref{eq:kappa}, which is equivalent to confinement pressure.  

No damping mechanism is explicitly considered at the macroscale by the DD solver, doing so would require computing evolving free energies of each micro ensemble and each load level and adopting an incremental approach in the algorithm, a feasible-yet-involved task (see \cite{Kostas_1}) that will be addressed in future work. 
The contact viscosity that is present at the micro level controls the relaxation time of the granular system, but does not provide a macro dissipation mechanism. 
Then, a natural question we would like to address pertains to the finiteness of the amplification in the DD results for all base load frequency values: the rigid bedrock assumption allows no radiational damping \cite{Kramer}, so the linear-elastic theory predicts infinite amplification in this setting if no material damping is present. This is due to unphysical ``division-by-zero'' mathematical artifacts that cannot happen in our time-domain simulations as we record the maximum amplitude once a steady-state has been attained.

Therefore one future work direction is bringing energy dissipation at the material level into the picture, what would be done by adding intergranular friction to the DEM microstructure model and following the pathway outlined by Karapiperis et al. \cite{Kostas_1}. This should enable the reproduction of the material damping associated to hysteresis loops seen in experiments \cite{hardin}.
Conversely, adding the radiational damping associated to a non-rigid interface between soil and base rock is more of a computational problem and neither a matter of material modeling \cite{Wolf} nor an immediate concern as we focus on soft soil deposits. Other directions are to include material rate-dependence and compare to low-intensity field records where the soil conditions are well-documented \cite{Kaklamanos,Thompson-Baise}.      

Let us highlight, finally, that we have verified our DD framework using (1) FEA with a strain-independent piece-wise approximation to the shear modulus in \Cref{Sec:DD-FEM}, and (2) an analytical solution using a linear fit for the shear modulus (see \Cref{Sec:TFs}). However, the logical path would have been the other way around: using the more realistic material response embedded in the DDCM procedure to validate simpler numerical approximations (FEA that does not account for non-linear soil response in the upper layers) and analytical formulae that assumes a strain-independent linear evolution of stiffness with depth. 

\section{Final remarks}

This text has introduced the novel multiscale data-driven paradigm to 1D site response analysis, and, as a prior step, it has also demonstrated the suitability of DDCM for solving time-domain wave propagation problems in 1D. The soil response datasets have been mined via microstructure RVE DEM simulations. 

A minimal model has been set up and validated via comparison to finite element analysis. It has been shown that the DD solution procedure can naturally handle elastic non-linear material behavior that would require complex modeling to have it accounted for in traditional FEA. The model has then been used to generate the base-to-top transfer function corresponding to an idealized soil column in a soil deposit resting on bedrock. The DD results resemble prior analytical results while displaying desirable high-frequency traits that are not captured easily by simplified constitutive laws.  

This paper lays the foundations of upcoming data-driven tools for both site response analyses and ground motion prediction. The next logical steps are to include material rate dependence and dissipation arising from intergranular friction. These have been left out of the scope of this text as the validation of a DDCM procedure of this case will previously require a thorough study of data-driven wave propagation and setting up complex simulations that include state-of-the-art constitutive models and direct comparison against costly full-resolution DEM simulations.

To conclude, let us make clear that in no way we are advocating for a conceptual primacy of DDCM over constitutive models anchored in plasticity theory; rather, one of the main arguments in favor of DDCM appeals to sheer convenience: phenomenological models require calibration of a sizeable number of parameters (e.g, 22 in Ref.~\cite{Boulanger}) and simulations require careful meshing and time-stepping to accurately resolve the plastic flows, while DDCM demands neither. 


\section*{Supplementary material}

The code necessary to reproduce the results presented in this paper can be retrieved from the second author's \textit{c4science} repository (\texttt{c4science.ch/source/DD\_1D-SRA}). A Mathematica notebook that solved the FEM equivalent problem presented in \Cref{Sec:DD-FEM} and evaluates the analytical transfer function in \Cref{Sec:TFs} can be obtained from the first author \textit{Github} (\texttt{github.com/jgarciasuarez}) in the repository \texttt{DD\_1D-SRA}.

\section*{Acknowledgments}
This work was supported in part by the Swiss National Science Foundation under the grant "Wear across scales" (200021\_197152). The authors thank Manon Voisin--Leprince for sharing her DEM acumen and for providing starter LAMMPS scripts.


\bibliography{references}
\bibliographystyle{acm}

\end{document}